a SpringerOpen Journal

RESEARCH  Open Access

# Image-guided therapy system for interstitial gynecologic brachytherapy in a multimodality operating suite

Jan Egger

**Abstract**

In this contribution, an image-guided therapy system supporting gynecologic radiation therapy is introduced. The overall workflow of the presented system starts with the arrival of the patient and ends with follow-up examinations by imaging and a superimposed visualization of the modeled device from a PACS system. Thereby, the system covers all treatments stages (pre-, intra- and postoperative) and has been designed and constructed by a computer scientist with feedback from an interdisciplinary team of physicians and engineers. This integrated medical system enables dispatch of diagnostic images directly after acquisition to a processing workstation that has an on-board 3D Computer Aided Design model of a medical device. Thus, allowing precise identification of catheter location in the 3D imaging model which later provides rapid feedback to the clinician regarding device location. Moreover, the system enables the ability to perform patient-specific pre-implant evaluation by assessing the placement of interstitial needles prior to an intervention via virtual template matching with a diagnostic scan.

## Introduction

With over eighty thousand new cases in 2010 and over Twenty-five Thousand deaths per year, gynecologic malignancies – including cervical, endometrial, and vaginal/vulvar cancers – are the 4th leading cause of death in women in the United States (Cancer Facts & Figures, American Cancer Society 2010). In general, therapy consists of three components: concurrent chemotherapy and external beam radiation followed by brachytherapy. During the external beam radiation stage a machine targets radiation beams at the pelvis from outside the body. In contrast to this, radioactive sources that deliver very high doses of radiation are placed directly inside the cancerous tissue during the brachytherapy stage. This is done by placing an applicator in the vaginal canal of the patient. In the past few years, several gynecologic cancer brachytherapy centers around the world have shown the benefit of using magnetic resonance imaging (MRI) scans to guide brachytherapy planning and incorporated it into their clinical practice. Viswanathan et al. (2007) was among the first to demonstrate that MRI significantly increased the coverage of the tumor by radiation dose as compared to computed tomography (CT).

Both, the European Society for Therapeutic Radiology and Oncology (ESTRO) and the American Brachytherapy Society (ABS) have recommended that T2-weighted MRI should be used for organs at risk (OARs) delineation and targeting in image-based cervical cancer brachytherapy. For a comprehensive explanation the reader is referred to a recent textbook on gynecologic radiation therapy and its references (Viswanathan et al. 2011) at this point, where the dosimetric and clinical gains from using MRI versus CT or ultrasound (US) are described in detail. Summarized, the ability to more accurately delineate the tumor and the surrounding normal tissue is the primary benefit in using MR compared to the more standard practice of CT. This leads to better dose escalation to the target volume while respecting dose constraints for the surrounding OARs. Studies have shown that CT may fail to distinguish cervical tumor from surrounding normal tissues such as small bowel, while MR determines the size, location, and paracervical involvement of the tumor and its relations to the applicator. The clinical practice of brachytherapy is well characterized using five components: 1) Applicator Choice and Insertion Techniques 2) Imaging Protocol

Correspondence: egger@uni-marburg.de
Department of Medicine, University Hospital of Giessen and Marburg (UKGM), Baldingerstraße, Marburg 35043, Germany

3) Contouring Protocol 4) Treatment Planning 5) Dose and Fractionation. Details of each step are also provided in the textbook of Viswanathan et al. (2011) and its references.

The purpose of this study is to introduce a system for supporting MR- and CT-guided gynecologic radiation therapy that covers all treatment stages (pre-, intra- and postoperative) with a focus on the five brachytherapy components introduced in the previous paragraph. To the best of the author's knowledge such a system has not yet been described. The system has been worked out hand-in-hand within an interdisciplinary team of physicians, computer scientists and engineers. Research highlights include linking a diagnostic imaging set in real-time to a 3D CAD model of a medical device and precise identification of catheter location in the 3D imaging model with real-time imaging feedback. Moreover, the system enables the ability to perform patient-specific pre-implant evaluation by virtually assessing the placement of interstitial needles prior to an intervention via template matching with a diagnostic scan.

The contribution is organized as follows: Materials and Methods section presents the materials and the methods, Results section presents the results, and Conclusions section concludes the contribution and outlines areas for future work.

## Materials and methods
### Equipment
The Brigham and Women's (BWH) Hospital in Boston has recently built an Advanced Multimodality Image-Guided Operating (AMIGO) suite, which enables 3 Tesla (3 T) MRI and PET/CT imaging during therapy (Egger et al. 2012a). AMIGO is an extension of the BWH's Image Guided Therapy (IGT) program, established in 1991 by Dr. Ferenc Jolesz. Launched in 2011 as the successor to the 0.5 Tesla "double-donut" Signa SP (General Electric Healthcare) interventional suite in which BWH teams performed over 3,000 surgical and interventional procedures, the charter of AMIGO is to continue these pioneering efforts with multimodal image guidance (Kapur et al. 2012) and intraoperative gynecological MRI data from AMIGO is freely available for download (Egger et al. 2012b).

### System
The overall architecture of the medical system is shown in Figure 1 and has been preregistered as an invention disclosure (Egger et al. 2012c) and was adapted from a stent simulation system (Egger et al. 2009). The workflow starts with the arrival and imaging of the patient in the preoperative stage. The preoperative images are used to make a diagnosis and measure the relevant sizes for gynecologic radiation therapy. This also leads to the decision if the patient is eligible for an intervention or not.

If the patient is eligible, an automatic request is send to the manufacturer and the inventory database and a first (medical) device selection is made. This selected device (e.g. Tandem and Ring/Ovoid +/−, interstitial needles or Vienna Applicator, interstitial needles alone) is modeled and visualized in the preoperative images from the patient and reviewed by a physician. Thereby, the system enables the virtual modeling and visualization of several instruments in the preoperative images of the patient, which allows a direct comparison of different devices to find the optimal one for treatment and dose delivery. Afterwards the selected device is ordered from the inventory and the modeled device (for example a 3D CAD model) is stored in a Picture Archiving Communication Software (PACS) system for the intra- and postoperative stage. In the intraoperative stage the patient is imaged with an intraoperative MRI (iMRI) and the modeled device from the PACS system is used for guidance during the intervention. Thus, the device is visualized in the patient's dataset to display the optimal position for dose delivery to the physician. The system ends in the postoperative stage with a follow-up examination by imaging and a superimposed visualization of the virtually modeled device from the PACS system.

The following sections describe the key components of the presented system in more detail, which include: the *Imaging* protocols that have been used to acquire the patient data, the *Linking* algorithm to register the 3D *Models* of the medical devices to the pre- and intraoperative patient datasets, the *Transfer* protocol to communicate between the different parts of the system, the *Contouring* of the patient's Organs at Risk and the *Planning* process for a precise dose distribution.

### Imaging
After applicator placement, T1- and T2-weighted MR images are acquired with a 3 Tesla scanner. Therefore, T2-weighted sequences – Fast Spin Echo (FSE), Turbo Spin Echo (TSE) – are performed in the axial, sagittal and coronal orientation. Additionally, paraaxial, parasagittal, and paracoronal slices oriented orthogonal and parallel to the applicator axis are taken. Hereby, the axial images are obtained from above the uterine fundus down to the inferior border of the symphysis pubis, and in case of any vaginal tumor extension even below this. The sagittal images on the other side are obtained between internal obturator muscles. Finally, the (para) coronal and paraxial images are obtained so they cover the tumor, the entire cervix, the corpus uteri, the parametria, and the vagina. The slice thickness for the T1-weighted MR images is 3 mm with no gaps and the slice thickness for the T2-weighted MR images is 4-5 mm with 1 mm or no gaps (Viswanathan et al. 2011, Kapur 2012).

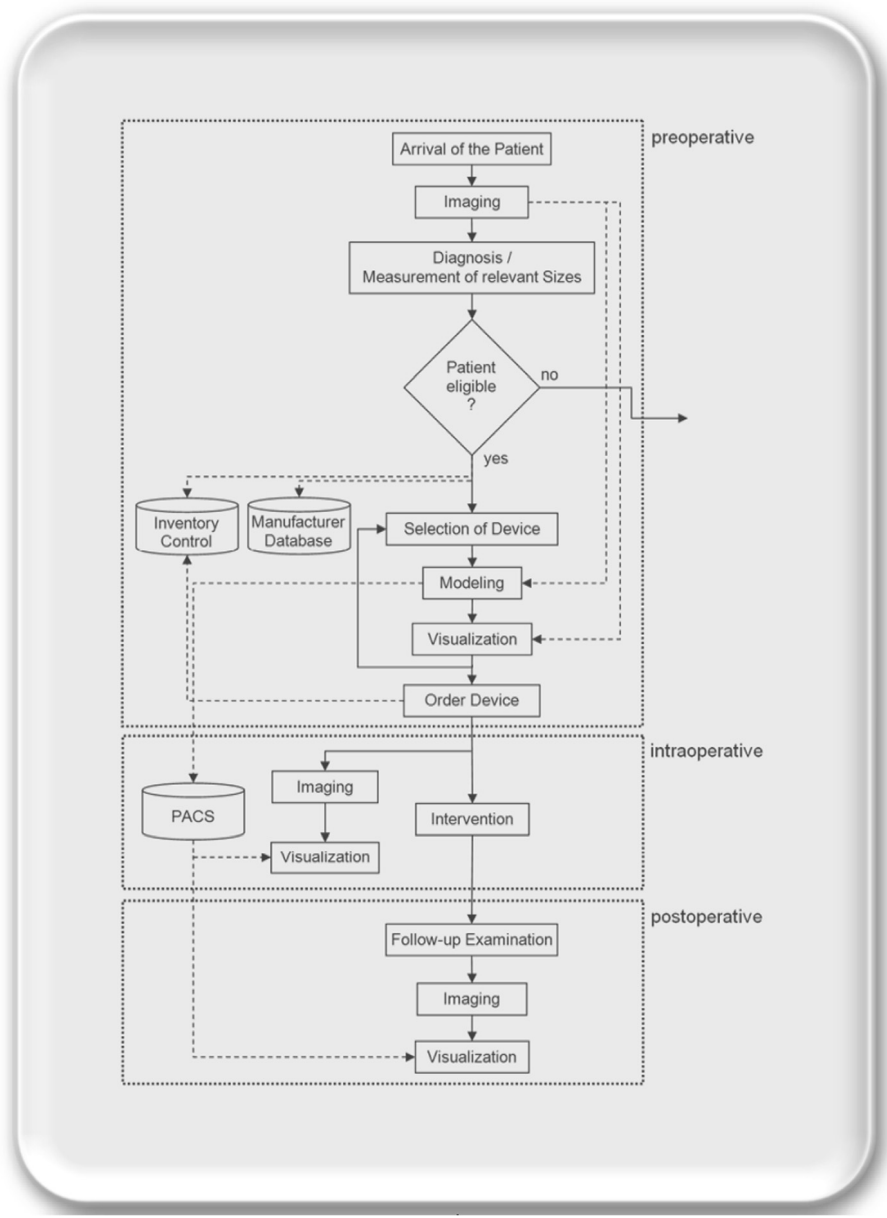

**Figure 1 The overall end-to-end workflow of the presented system, starting with the arrival of the patient and finishing with follow-up examinations.**

### Linking

For linking a diagnostic and/or intraoperative imaging set to a 3D CAD model of a medical device a software prototype called iGyne has been developed, which allows identification of catheter location in the 3D imaging model with real-time imaging feedback. The first version of iGyne has been implemented in the medical platform MeVisLab (www.mevislab.de) as an own C++ module, end of 2011 (Egger et al. 2012a, Kapur et al. 2012). In the meantime there exist public available software modules for 3D Slicer (3DSlicer 2013, Pieper et al. 2004, Fedorov et al. 2012). 3D Slicer – or Slicer – is a free and open source software package for visualization and image analysis primarily used in the medical domain and has been developed by the Surgical Planning Laboratory (SPL) of the Brigham and Women's Hospital in Boston.

The intraoperative usage of iGyne in AMIGO consists of several steps (Chen et al. 2012): in a first step the CAD models of the medical devices are loaded into the software. Then, the patients MR scan with the template sutured to the perineum and the obturator placed in the vaginal canal is transferred via the DICOM protocol to iGyne. Next, an initial rigid registration is computed over three corresponding point pairs manually provided by the user, which is afterwards refined using the Iterated Closest Point algorithm for rigid registration (Besl and McKay 1992, Xiaojun et al. 2007). Finally, iGyne enables the rendering of virtual needles in the 2D and 3D views which allows the visualization and observation of spatial relationships among the needles, tumors, and surrounding anatomical structures. Thus, supporting the determination of the amount of needles and their positions as well as their insertion depths for a patient.

### Models
The 3D models of the medical devices like the interstitial template and vaginal obturator (Figure 2) were created in advance using CAD software (SolidWorks, Dassault Systèmes SolidWorks Corp., MA). All models were reveres-engineered by measuring the detailed dimensions from the clinically used devices. Afterwards, the models were converted to an industry standard format (STL).

### Transfer
An important factor for a medical system using software and hardware from different manufactures is the data exchange. To have the intraoperative patient data immediately available in iGyne the DICOM listener from Slicer has been used. Thus, the patient acquisitions could be pushed automatically from the scanner to iGyne for further processing (e.g. *Linking* and *Contouring*). However, for an intraoperative navigation or guidance of catheter placement a more sophisticated solution is needed, where a navigation system or ultrasound is integrated into the whole system. This can be realized via the OpenIGTLink network protocol developed at the BWH (Tokuda et al. 2010). OpenIGTLink is a new, open, simple and extensible peer-to-peer network communication protocol for IGT which has already been used successfully for prostate interventions (Tokuda et al. 2008). The protocol provides a standardized mechanism to connect hardware and software by the transfer of coordinate transforms, images, and status messages. The advantage of OpenIGTLink is its simple specification, initially developed through a collaboration of academic, clinical and industrial partners for developing an integrated robotic system for MRI-guided prostate interventions (Fischer et al. 2008). It was designed for use in the application layer on the TCP/IP stack, allowing researchers to develop a prototype system that integrates multiple medical devices using the standard network infrastructure (Egger et al. 2012d). The OpenIGTLink protocol itself does not include mechanisms to establish and manage a session. It only defines a set of messages, which is the minimum data unit of this protocol. In Summary, an OpenIGTLink message contains all information necessary for interpretation by the receiver and begins with a 58-byte header section, which is common to all types of data, followed by a body section. The format of the body section varies by the data type that is specified in the header section. Since any compatible receiver can interpret the header section, which contains the size and the data type of the body, every receiver can gracefully handle any message, even those with an unknown data type, by ignoring incompatible messages without the system crashing.

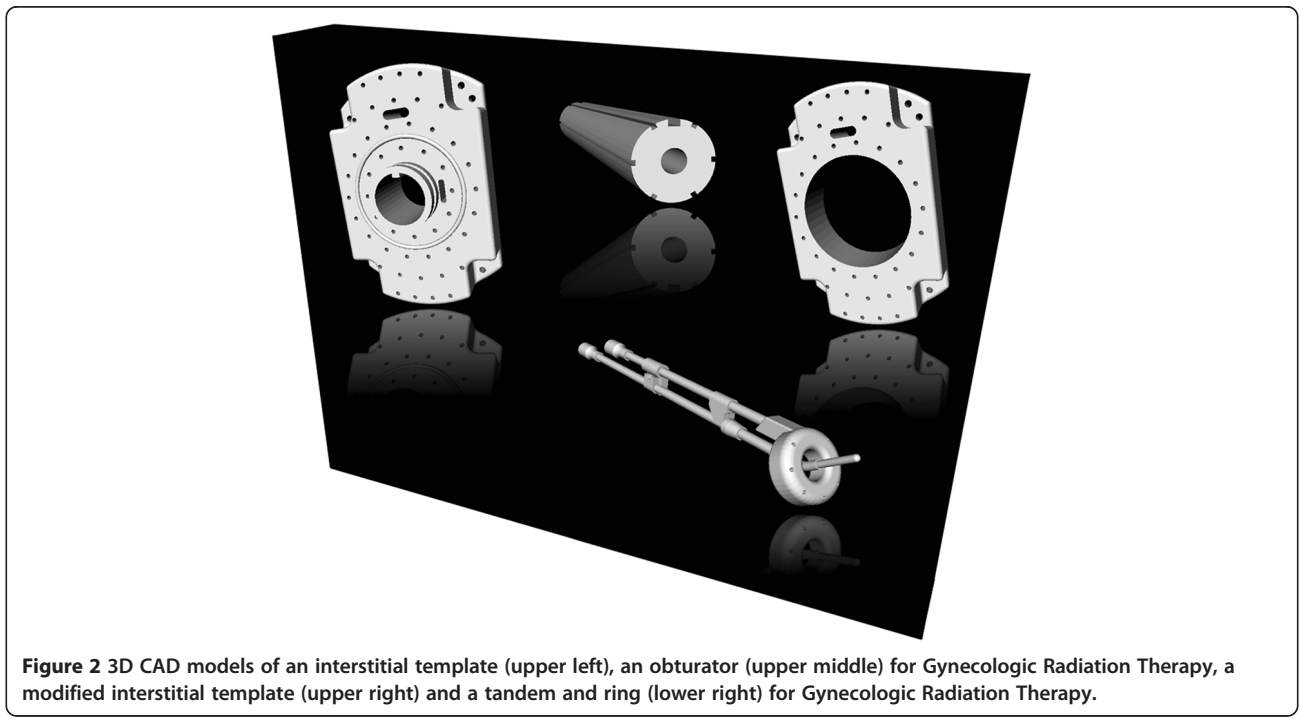

**Figure 2** 3D CAD models of an interstitial template (upper left), an obturator (upper middle) for Gynecologic Radiation Therapy, a modified interstitial template (upper right) and a tandem and ring (lower right) for Gynecologic Radiation Therapy.

### Contouring

According to the GYN GEC ESTRO guidelines (Haie-Meder et al. 2005, Pötter et al. 2006), three types of structures are outlined in the patient data for the proposed system: the gross tumor volume (GTV), the clinical tumor volume (CTV) and the organs at risk (OAR). Hereby, the GTV is the tumor – as seen by the physician in the patient's images and clinical exams – which gets the maximum radiation dose for treatment. For the CTV two areas are defined: the intermediate risk clinical tumor volume (IR-CTV) and the high-risk clinical tumor volume (HR-CTV). Thereby, the HR-CTV is the GTV including the cervix and any suspicious areas where disease remains at the time of brachytherapy, like a uterine invasion or parametrial involved tissues. The IR-CTV is an extension of the HR-CTV by 1 cm and any disease extension at the time of diagnosis. Finally, the OARs consist of the bladder, the rectum, the sigmoid and the small bowel adjacent to the uterus which should get a minimal possible radiation dose (Viswanathan et al. 2011, Kapur 2012). For the segmentation of the pelvic structures for gynecologic brachytherapy we studied the capabilities available in Slicer. As a result, segmentation of the bladder could be achieved accurately using the implementation of the GrowCut algorithm in Slicer. But, manual segmentation was required to achieve segmentation results for the tumor and the rectosigmoid (Egger et al. 2012e).

### Planning

The treatment planning of the presented system can be performed by using commercial software packages from Nucletron (Plato, OncentraGYN) and Varian (BrachyVision). These software applications work with different types of image modalities like MR, CT, X-ray, Ultrasound, and therefore a suitable for a multimodal operation suite like *AMIGO*. In a first step, the geometry of the inserted applicator is reconstructed, which is done by using its appearance in the images (Pernelle et al. 2013), followed by the generation of a treatment plan based on the prescribed doses to the contours GTV, CTV, and constraints on the OAR. Thereafter, manual refinement is performed to obtain optimal dose distribution covering the target (tumor). Thus, the isodose distribution is evaluated after each modification, which includes a detailed analysis of the dose-volume parameters (DVH) for CTVs and OARs (note: during the external beam radiation therapy (EBRT) the patients receive a dose of 40–50 Gy and the following brachytherapy provides an extra 3–5 fractions of 5.5-7 Gy each. The overall aim is to deliver a dose of 80–90 Gy to the high-risk clinical tumor volume and the dose to the OAR is assessed through measurement of dose rate. This measurement can be achieved with one of the following two methods: (1) the common ICRU 38 reference points, or more recently, (2) 3D volumes like the D2cc and D0.1 cc, which are defined for the bladder, the rectum/sigmoid, and small bowel. The D2cc tolerances are assumed to be approximately 90 Gy for the bladder, 70 Gy for rectum/sigmoid, and 55 Gy for the small bowel (Viswanathan et al. 2011, Kapur 2012).

### Results

In this contribution an overall system for supporting Gynecologic Radiation Therapy with focus on a multi-modality operating suite, like AMIGO, has been worked out. The Image-Guided Therapy system starts with the patient arrival and imaging. Then, diagnosis and measurement of relevant sizes for intervention in a multi-modality operating suite are made leading to automatic inventory control and manufacturer request. Next, a first device is selected (e.g. Tandem and Ring/Ovoid +/– interstitial needles or interstitial needles alone) that is modeled in the preoperative images. Thereby, virtual modeling and visualization of several instruments for direct device comparison is enabled to find the optimal one. Afterwards, the device is ordered from the hospital inventory and the modeled device is stored (PACS). In the intraoperative stage the patient is imaged and the stored device allows guidance to an optimal position for the subsequent dose delivery. The system ends with follow-up examinations by imaging and a superimposed visualization of the modeled device from the PACS. In order to increase the physician's speed and monitor the consequences of inserting interstitial catheters in real-time a software prototype has been developed. The prototype has been realized as an own module called iGyne (interstitial Gynecologic Radiation) under the medical research platform 3D Slicer in C++:

https://github.com/xjchen/igyne

and Python:

https://github.com/gpernelle/iGynePy
http://www.slicer.org/slicerWiki/index.php/Documentation/Nightly/Extensions/iGyne

Figure 3 shows several screenshots of the iGyne module under Slicer. The screenshots show the 3D CAD model of the interstitial template that has been fitted to MRI scans. On the left side of the iGyne interface an interstitial planning sheet is provided which allows virtual preplanning of single needles (depth and length). The software also enables rendering of the planned interstitial needles in different 2D slices. The iGyne Slicer module could be performed the planning in *real-time* on a

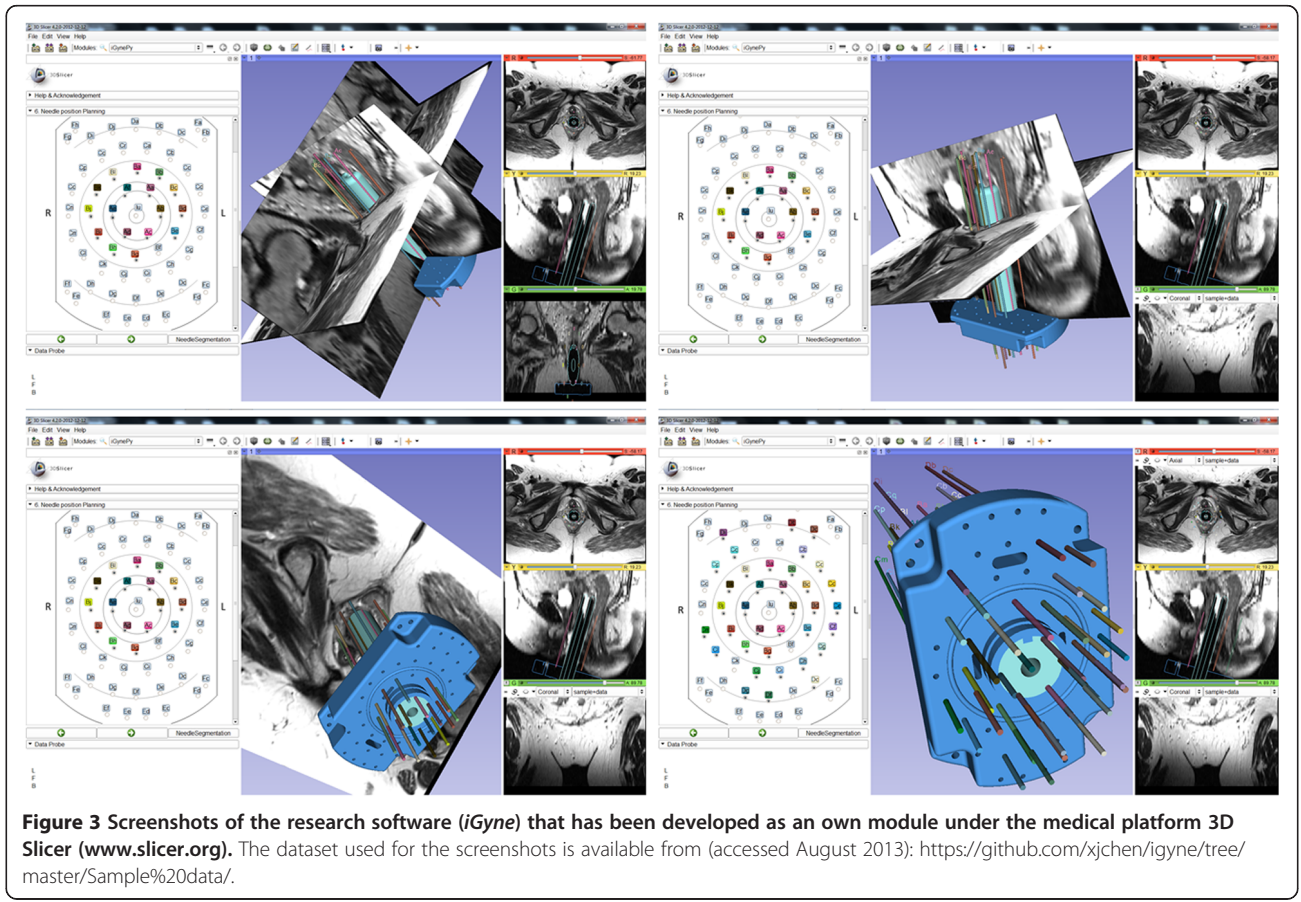

**Figure 3 Screenshots of the research software (*iGyne*) that has been developed as an own module under the medical platform 3D Slicer (www.slicer.org).** The dataset used for the screenshots is available from (accessed August 2013): https://github.com/xjchen/igyne/tree/master/Sample%20data/.

Laptop with Intel Core i5-2520 M CPU, 2 × 2.5 GHz, 4 GB RAM, Windows 7 Version, Service Pack 1, 32Bit.

## Conclusions

In this contribution, an overall Image-Guided Therapy system for supporting Gynecologic Radiation Therapy has been introduced that covers all therapy stages from patient arrival/diagnosis (preoperative), intervention (intraoperative) to follow-up examinations (postoperative). To the best of the author's knowledge, this is the first time an overall medical system for Image-Guided Therapy supporting Gynecologic Radiation Therapy has been introduced. In summary, the achieved research highlights of the presented work include:

- linking a diagnostic imaging set in real-time to a 3D CAD model of a medical device;
- precise identification of catheter location in the 3D imaging model with real-time imaging feedback and
- ability to perform patient-specific pre-implant evaluation by assessing in the computer the placement of interstitial needles prior to an intervention via virtual template matching with a diagnostic scan.

The overall workflow of the designed system has been described in detail step-by-step: starting with the arrival of the patient and finishing with follow-up examinations.

However, there are several areas of future work, like the integration of intraoperative navigation (e.g. intraoperative ultrasound (iUS), electromagnetic (EM) tracking or optical navigation) into the system (Wang et al. 2013). In contrast to prostate interventions where navigation systems have been used (Tokuda et al. 2008, 2010) medical navigation systems have not yet found their way into gynecological interventions. However, there are other research studies that have influence on a medical navigation system for gynecologic radiation therapy, like the exploration of female patient positions undergoing gynecological surgeries (Power 2003). Moreover, a motion study between supine and lithotomy positions of the female pelvis should be performed by acquiring magnetic resonance imaging data in different stirrups adjustments. Critical is the deformation that has been studied for the prostate (Hirose et al. 2002, Bharatha et al. 2001) and also in particular for prostate brachytherapy (Bellon et al. 1999) but to the best of the author's knowledge not for the female pelvis.

**Competing interests**
The author in this paper has no potential competing interests.

**Authors' information**
egger@uni-marburg.de
egger@med.uni-marburg.de
egger@staff.uni-marburg.de

**Acknowledgements**
First of all, the author would like to acknowledge the *AMIGO* team in enabling this study and the members of the Slicer community for their support. In addition, the author wants to acknowledge Xiaojun Chen, Ph.D. and Guillaume Pernelle for implementing the C++ and Python versions of *iGyne* under *Slicer*. Moreover, the author wants to thank Sam Song, Ph.D. for generating the CAD models of gynecologic devices and Ehud Schmidt, Ph.D. for proofreading. Furthermore, the author thanks Professor Dr. Horst K. Hahn and Fraunhofer MeVis in Bremen, Germany, for their support by providing an academic license for the MeVisLab software.

Received: 31 May 2013 Accepted: 19 August 2013
Published: 21 August 2013